\documentclass[twocolumn,floatfix,pre,showpacs]{revtex4-1}
\usepackage{graphicx}
\usepackage{amssymb}
\usepackage{amsmath}
\usepackage{color}
\usepackage{psfrag}
\usepackage{epsfig}
\usepackage{bbm}
\usepackage{bm}
\usepackage{dsfont}

\newcommand{\rb}[1]{\left( #1 \right)}

\newcommand{\ew}[1]{\langle #1 \rangle}
\newcommand{\beq}{\begin{eqnarray}}
\newcommand{\eeq}{\end{eqnarray}}

\newcommand{\eq}[1]{Eq.~(\ref{#1})}
\newcommand{\fig}[1]{Fig.~\ref{#1}}

\newcommand{\kett}[1]{| #1 \rangle\!\rangle }
\newcommand{\braa}[1]{\langle\!\langle #1|}
\newcommand{\eww}[1]{\langle\! \langle #1\rangle\! \rangle}

\begin{document}
\title{A minimal model for short-time diffusion in periodic potentials}
\author{Clive Emary, Robert Gernert, and Sabine H.~L.~Klapp}
\affiliation{
  Institut f\"ur Theoretische Physik,
  Hardenbergstra\ss{}e 36,
  Technische Universit\"at Berlin,
  D-10623 Berlin,
  Germany
}

\date{\today}
\begin{abstract}
  We investigate the dynamics of a single, overdamped colloidal particle, which is driven by a constant force
  through a one-dimensional periodic potential. We focus on systems with large barrier heights where the lowest-order cumulants of the density field, that is, average position and the mean-squared displacement, show nontrivial (nondiffusive) short-time behavior characterized by the appearance of plateaus. 
  We demonstrate that this ``cage-like'' dynamics can be well described by a discretized master equation model involving two states (related to two positions) within each potential valley. 
  Nontrivial predictions of our approach include analytic expressions for the plateau heights and an estimate of the ``de-caging time'' obtained from the  study of deviations from Gaussian behaviour.
  The simplicity of our approach means that it offers a minimal model to describe the short-time behavior of systems with hindered dynamics.
\end{abstract}
\maketitle

\section{Introduction \label{SEC:INTRO}}

\begin{figure}[tb]
  
  \includegraphics[width=0.9\columnwidth,clip]{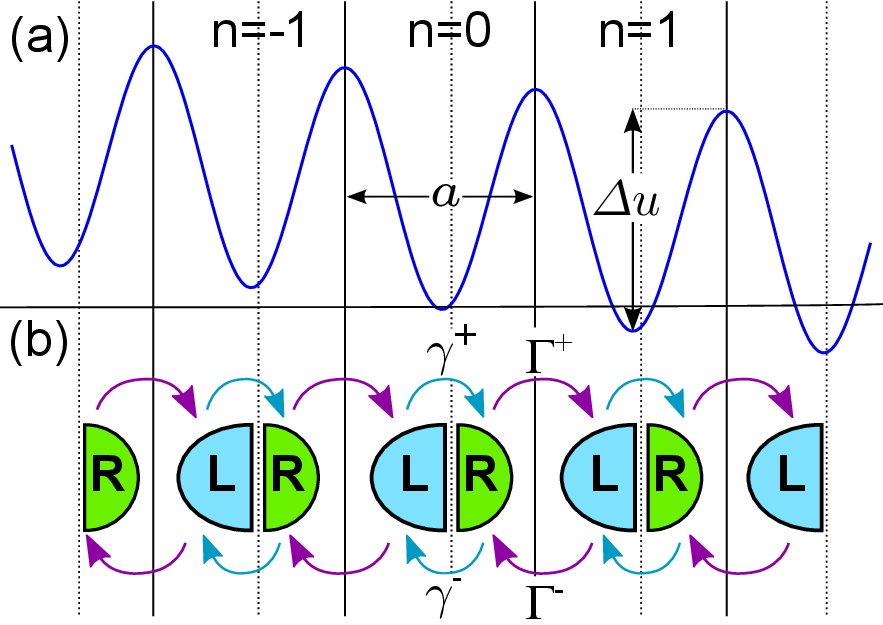}
  \caption{ 
    (Color online)
    (a) The potential landscape of a particle in a tilted washboard.  The period is $a$ and the minimum peak-to-trough height is $\Delta u$.
    (b) The discrete two-state-per-well (2SW) model in which the state of the system is described by well index $n$ and $\alpha=L,R$, describing whether the particle is found in the left or right side of the well.  The particle therefore assumes discrete positions at $x=n a + x_{L/R}$, where $x_\alpha$ is the displacement of internal state $\alpha$.
    Transitions between the states are described by a rate equation with rates $\gamma^\pm$ and $\Gamma^\pm$ as shown.
    \label{FIG:SKETCH}
  }
\end{figure}

The transport of particles in modulated potentials plays a fundamental role in diverse fields such as colloidal particles on topologically or energetically 
structured surfaces \cite{Tierno10,Dalle11}, particles in optical lattices \cite{Siler10} and optical line traps \cite{Lopez08}, 
biased Josephson junctions \cite{Kivioja05}, and in biophysical processes \cite{Ros05,Hanggi09}. 
In many of these cases, the dynamics can be described
as overdamped (i.e., noninertial) Brownian motion in one spatial direction. A generic model in this context is a particle in a periodic ``washboard" (e.g., sinusoidal) potential,
which is tilted by the constant biasing force, see \fig{FIG:SKETCH}(a). The resulting particle current in the long-time limit was calculated analytically more than 50 years ago by Stratonovich \cite{Stratonovich58,Risken}. More recently,
Reimann {\it et al.} \cite{Reimann01} 
derived an analytical solution for the corresponding diffusion coefficient related to the long-time limit of the mean-squared displacement,
$\langle \Delta x^2(t)\rangle$. This diffusion coefficient in the modulated potential can become much {\it larger} than that in a free system, 
contrary to the intuitive idea that confinement slows down diffusion. This ``giant diffusion" effect has also been found in experiments \cite{Lee06,evsti08,Mu09}.
Taken together, one can state that the long-time properties of particles in periodic forces are nowadays well understood.

Much less is known on the {\it short-time} dynamics of driven (or undriven) overdamped particles in periodic forces \cite{Lindenberg07}.
However, it is precisely this regime which has recently attracted significant theoretical and experimental interest (see, e.g.\ Refs.\ \cite{Vorse07,Dalle11,Hanes12,Wang12}). 
One reason for this interest is the similarity of the observed short-time dynamics with that of 
systems approaching a glass transition \cite{Vorse07,Schweizer05,Doliwa03,Odagaki90,Goychuk11} or, more generally,
a state characterized by complex energy landscapes yielding trapping. A key quantity in this context is the mean-squared displacement as function of time. For (noninteracting and interacting) colloids in modulated potentials, this quantity has already been investigated by computer simulations of the corresponding Langevin equations \cite{Vorse07},
numerical solutions of the Fokker-Planck (Smoluchowski) equation \cite{lichtner10}, 
as well as in experiments \cite{Dalle11}. Provided that the barrier height is sufficiently large, $\langle \Delta x^2(t)\rangle$
typically grows linearly with time initially, reflecting diffusive motion within one valley. At intermediate time a plateau appears signaling that the particles are trapped (subdiffusive behavior). Only at long times, do the particles
eventually overcome the barrier, and $\langle \Delta x^2(t)\rangle$ approaches linear time-dependence (these three dynamic regimes can be seen in \fig{FIG:C2t}).
These dynamic features of a single particle in a modulated potential resemble the single-particle dynamics of various many-particle systems in kinetically trapped or glassy states \cite{Chaud99,Chaud07}, examples being
dipolar particles trapped in chains \cite{Jordanovic11}, colloidal gels \cite{Gado07}, and (binary) Lennard-Jones glasses \cite{Chaud99}.
Indeed, from the perspective of glasses (or gels), the modulated one-particle potential is thought to arise from interaction effects, that is, the formation of cages formed by the neighboring particles\cite{Vorse07}.
Hopping over a barrier then corresponds to the escape from the cage.

Contrary to the long-time properties discussed above there is, so far, no simple analytical approach towards the transport properties in the short-time regime. 
Attempts have been made to derive, on an approximate level, explicit formulae predicting the height of the plateau, as well as the crossover time from the sub-diffusive into the diffusive regime \cite{Vorse07,Dalle11}. This yields good estimates, but is clearly far away from detailed, quantitative predictions. To overcome this gap, we propose in the present paper 
a discretized model, which we treat by the master equation formalism. 
Clearly, the investigation of diffusion problems via discretized models and master equations has a long tradition \cite{Haus87}.
In particular, it is well known that conventional master equation approaches without memory effects yield a linear time dependence of all cumulants, i.e., normal diffusive behavior. 
More complex dynamics, also resembling that of glassy systems, is displayed by models based on continuous time random walks \cite{Montroll65,Heuer08,Chaud08,Odagaki90,Doliwa03}
and fractional Brownian motion \cite{Sokolov05,Goychuk11}

Here we introduce a discrete model in which the particle can adopt one of two positions (``states'') within each valley. Transitions both between these states and between wells are described by rates.
Using techniques similar to those used in, e.g., the theory of full counting statistics in quantum transport \cite{bag03,fli04,Emary2009}, we calculate the cumulant generating function for the particles position within this rate equation approach.  The treatment is simple, requiring just manipulation of  two-by-two matrices, and analytic results can be obtained.
In parallel, we calculate the density distribution and the cumulants numerically from the Smoluchowski equation (SE) and demonstrate that our ``two-state-per-well'' (2SW) model gives an excellent account of the short-time dynamics of interest here.  In particular, we obtain analytic expressions for the plateau heights, as well as estimates of the ``de-caging'' or ``cross-over'' time from the study of the non-Gaussian parameter, the kurtosis.

The paper is structured as follows. In Sec.\ \ref{SEC:SE} we describe the full Smoluchowski equation with a periodic potential and in Sec.\ \ref{SEC:TS/W} we present our rate equation model.  The three dynamical regimes 
are investigated in Sec.\ \ref{SEC:LIM} and in Sec.\ \ref{SEC:RATES} we describe the microscopic evaluation of the model parameters.  Section \ref{SEC:RESULTS} compares the results between model and numerically exact results. Finally, in Sec.\ \ref{SEC:CONCS}, we conclude.

\section{The Smoluchowski equation with a periodic potential \label{SEC:SE}}

We are interested in the diffusion of a particle subjected to a periodic force in one dimension.  For our purposes the full description of the system is given by the probability distribution function $\rho(x,t)$, which we take to be governed by the Smoluchowski (overdamped Fokker-Planck) equation \cite{Risken} 
\beq
  \dot{\rho}(x,t) = \mathcal{L}\rho(x,t)
  \label{SE}
  ,
\eeq
with the differential operator
\beq
  \mathcal{L} = 
  \frac{D_0}{kT} \left[
    kT \partial^2_x + u''(x) + u'(x) \partial_x
  \right]
  \label{superL}
\eeq
where $kT$ is the thermal energy, $D_0$ the so-called short-time diffusion coefficient, and $u(x)$ the modulated potential.

In this work we shall consider as example a tilted washboard potential (our approach is general, however).  The potential reads
\beq
  u(x) = u_0 \sin^2 (\frac{\pi x}{a}) - F x
  ,
\eeq
with $a$ the period of the potential, and parameters $u_0$ and $F$ describe the depth and tilt of the potential respectively.  A section of this potential is plotted in \fig{FIG:SKETCH}(a).

We assume that the form of the potential is such that it may be meaningfully divided up into a number of wells, defined between subsequent maxima, and labelled $n = \ldots,-2,-1,0,1,2,\ldots$.
This occurs when the difference between the potential maxima and maxima, $\Delta u$,
is large compared with the thermal energy of the particles: $\Delta u \gg kT$.

The main quantities of interest here are the cumulants of the particles position, which we denote $\ew{x^k}_c$ for the $k$th cumulant.  Of particular interest is the second cumulant or variance $\ew{x^2}_c = \ew{\Delta x^2} = \ew{x^2}-\ew{x}^2$.  We will also consider the kurtosis
\beq
  \kappa(t)
  =
  \frac{\ew{x^4(t)}_c}{\ew{x^2(t)}_c^2}
  \label{kappa}
\eeq
as a measure of the deviations from a Gaussian distribution.  We note that, in a single particle system, the kurtosis is trivially related to the $\alpha$-parameter often studied in the context of glass formation (see, e.g.\ Ref.~\cite{Vorse07}): $\alpha(t) =  \frac{1}{3}\kappa(t)$.

\section{Two-state-per-well model \label{SEC:TS/W}}

Our aim is to model the SE for the probability density defined as a function of continuous variable $x$ with a simpler model involving probabilities for finding the particle in one of a set of discrete states, defined by regions of the $x$ axis.
The most obvious approach of this type would be to associate one state with each well of the potential and write down a rate-equation for the probability $p^{(n)}(t)$ of finding the particle in well $n$ at time $t$.  This one-state-per-well approach is discussed in Appendix~\ref{APP:1S/W} and it is clear that all position cumulants are linear in time for all time.  Such a simple rate equation is therefore inadequate to describe the short-time dynamics of interest here.

The model that we shall focus on is the next most complicated model in this family, and possesses two states per well (2SW), which correspond to whether the particle is to be found in the left or right side of the well, see Fig.~\ref{FIG:SKETCH}(b).  In this model, the location of the particle is described by two discrete indices: the integer well index, $n$, and the internal index $\alpha = L,R$ and $p_\alpha^{(n)}$ is the corresponding probability.  For calculating the position cumulants, we take the particles to be localised at positions $x = n a + x_{L/R}$, where $x_L$ and $x_R$ are the offsets of the left and right states respectively.

Let us assume that transitions between these states can be described by rates and define rates $\gamma^+$  and  $\gamma^-$ to describe hopping within a single well, and rates $\Gamma^+$  and $\Gamma^-$ to describe hopping between the wells. In both cases, superscript $+$ indicates movement to the right and $-$, movement to the left.  The rates are homogeneous due to the periodicity of the force and, since we assume that the barriers are high ($\Delta u\gg kT$), the intra-well rates $\gamma^\pm$ are significantly larger than the inter-well rates $\Gamma^\pm$ \cite{FN}.  The system can then be described by the rate equation system
\begin{widetext}
\beq
  \frac{d}{dt} 
  \rb{
    \begin{array}{c}
      p^{(n)}_L \\
      p^{(n)}_R 
    \end{array}
  }
  &=&  
  \rb{
    \begin{array}{cc}
      -\gamma^+ -\Gamma^-& \gamma^- \\
      \gamma^+ & -\gamma^- -\Gamma^+
    \end{array}
  }
  \rb{
    \begin{array}{c}
      p^{(n)}_L \\
      p^{(n)}_R 
    \end{array}
  }
  +
  \rb{
    \begin{array}{cc}
      0 & \Gamma^+ \\
      0 & 0
    \end{array}
  }
  \rb{
    \begin{array}{c}
      p^{(n-1)}_L \\
      p^{(n-1)}_R 
    \end{array}
  }
  +
    \rb{
    \begin{array}{cc}
      0 & 0 \\
      \Gamma^- & 0
    \end{array}
  }
  \rb{
    \begin{array}{c}
      p^{(n+1)}_L \\
      p^{(n+1)}_R 
    \end{array}
  }
  \label{ME1}
  .
\eeq
\end{widetext}
Denoting the vector of probabilities with the notation $\kett{\rho^{(n)}}$, this master equation can be written in matrix form as \cite{bag03, fli04,Emary2009}
\beq
  \frac{d}{dt} \kett{\rho^{(n)}}
  =
 \mathcal{W}_0\kett{\rho^{(n)}} 
 +
 \mathcal{W}_+
 \kett{\rho^{(n-1)}}
 +
 \mathcal{W}_-
 \kett{\rho^{(n+1)}}
 \label{ME2}
 ,
\eeq
where $\mathcal{W}_{0,\pm}$ are two-dimensional matrices defined according to \eq{ME1}.  
We then define the Fourier-transformed vector $\kett{\rho(\chi;t)} = \sum_n e^{i n \chi}  \kett{\rho^{(n)}(t)}$, such that, by Fourier-transforming \eq{ME2}, we arrive at
\beq
  \frac{d}{dt} \kett{\rho(\chi;t)}
  =
  \mathcal{W}(\chi)\kett{\rho(\chi;t)}
  ,
\eeq
with
\beq
  \mathcal{W}(\chi) = 
    \mathcal{W}_0
    +
    e^{i\chi}
    \mathcal{W}_+
    +e^{-i\chi}
    \mathcal{W}_-
    .
\eeq
Assuming that the particle starts off in the $n=0$ well with distribution $\kett{\rho^{(0)}(0)} = \rb{p_L^{(0)}(0),p_R^{(0)}(0)}$, the solution of this equation is
\beq
  \kett{\rho(\chi;t)} = e^{ \mathcal{W}(\chi) t} \kett{\rho(0;0)} 
  .
\eeq

Based on this solution, the cumulant generating function (CGF) of the particle position is
\beq
  \mathcal{F}(\chi;t) = \log \braa{\widetilde{\phi}_0} e^{i \chi \Delta}  e^{ \mathcal{W}(\chi) t} \kett{\rho(0;0)} 
  \label{CGFx}
\eeq
with internal position matrix
\beq
   \Delta
   \equiv  
   \frac{1}{a}
   \rb{
    \begin{array}{cc}
      x_L & 0 \\
      0 & x_R
    \end{array}
  },
 \eeq
and where multiplication from the left with $\braa{\widetilde{\phi}_0} = (1,1)$  traces over internal states \cite{FNtracevec}.
Differentiation of the CGF w.r.t. $\chi$ and taking the $\chi\to0$ limit generates the cumulants:
\beq
  \ew{x^k(t)}_c = a^k \left.\frac{\partial^k}{\partial (i\chi)^k}  \mathcal{F}(\chi;t) \right|_{\chi=0}
  \label{cums}
  .
\eeq
A proof that the CGF of \eq{CGFx} does indeed give these cumulants is given in Appendix \ref{APP:CGF}. In the following we will always start the particle localised within a single well at either position $x_L$ or $x_R$.

\section{Limits \label{SEC:LIM}}
Calculating the CGF just involves manipulation of two-by-two matrices and various limiting cases can easily be extracted.  The three distinct phases of time evolution can be looked at in turn.

\subsection{Short times: linear growth}

Since the intra-well rates $\gamma^\pm$ are larger than the inter-well rates $\Gamma^\pm$, in the short-time limit all the dynamics occurs within the initial well. We therefore set $\Gamma^\pm=0$ in the CFG of \eq{CGFx} to obtain
\beq
  \mathcal{F}(\chi;t) = \log \braa{\widetilde{\phi}_0} e^{i \chi \Delta}  e^{ \mathcal{W}_\mathrm{intra}t} \kett{\rho(0;0)} 
 \label{CGFshort}
\eeq
with
\beq
  \mathcal{W}_\mathrm{intra}
  =
  \rb{
    \begin{array}{cc}
      -\gamma^+ & \gamma^- \\
      \gamma^+ & -\gamma^-
    \end{array}
  }.
\eeq
To assess the initial movement of the particle, we can expand this result to lowest order in $t$.
Of particular importance is the evolution of the first two cumulants from initially localised distributions at $x_{L/R}$. These read
\beq
  \ew{x(t)} &\sim& x_{L/R} \pm  \gamma^\pm (x_R - x_L) t;
  \nonumber\\
  \quad\ew{x^2(t)}_c &\sim&  \gamma^\pm(x_R - x_L)^2 t,
  \label{TSQlinear}
\eeq
where the upper- (lower-) sign corresponds to an initial position $x_L$ ($x_R$).

\subsection{Intermediate times: plateaus \label{SEC:LIMPLATEAU}}

Going beyond first order in time, \eq{CGFshort} predicts a first cumulant, for example, which reads
\beq
  \ew{x(t)}_c &=& \frac{\gamma^- x_L + \gamma^+ x_R}{\gamma^+ + \gamma^-}
  \nonumber\\
  &&
  +
  \frac{
    \gamma^+ p_L^{(0)}(0) - \gamma^-p_R^{(0)}(0)
    }{
      \gamma^+  + \gamma^-
    }
  (x_L-x_R)e^{-\gamma t}
  \label{x1c_2SWtime}
  .
\eeq  
The first cumulant therefore simply relaxes exponentially with rate $\gamma =  \gamma^+ + \gamma^-$ to an asymptotic value within the initial well.  The other cumulants behave similarly, with the expression for the $k$th cumulant being composed of a sum of terms with time-dependences $e^{-m \gamma t}$ where index $m$ is an integer with maximum value $k$. 
Asymptotically then, the cumulants within a single well reach values which correspond to the plateau values.
For the first four cumulants, the plateau values read
\beq
  \ew{x}^\mathrm{plat}_c &=& \frac{\gamma^+ x_L + \gamma^- x_R}{\gamma^+ + \gamma^-}
  \nonumber\\
  \ew{x^2}^\mathrm{plat}_c &=& \frac{\gamma^+  \gamma^- }{(\gamma^+ + \gamma^-)^2}(x_L-x_R)^2
  \label{plat1}\\
  \ew{x^3}^\mathrm{plat}_c &=& \frac{\gamma^+  \gamma^- (\gamma^+-\gamma^-)}{(\gamma^+ + \gamma^-)^3}(x_L-x_R)^3
  \nonumber\\
  \ew{x^4}^\mathrm{plat}_c &=& 
  \frac{\gamma^+  \gamma^- 
  ({\gamma^-}^2 -4\gamma^+\gamma^- +{\gamma^+}^2)}
  {(\gamma^+ + \gamma^-)^4}
  (x_L-x_R)^4\nonumber
  .
\eeq
These are the ``caged values''.  On time-scales larger than $\rb{\Gamma^{\pm}}^{-1}$, the particle begin to leave the first well and the values move away from the above.

\subsection{Long time-limit \label{SEC:LONGTIMELIMIT}}
In the asymptotic $t \to \infty $ limit, the complete CGF of \eq{CGFx} becomes
\beq
   \mathcal{F}^\infty_x(\chi;t) 
   \equiv 
   \lim_{t\to \infty}\mathcal{F}^\infty_x(\chi;t)
   =  
    \lambda_0(\chi)t
    \label{Flong}
\eeq
where $ \lambda_0(\chi)$ is that eigenvalue of $\mathcal{W}(\chi)$ which behaves as
$\lim_{\chi\to 0}\lambda_0(\chi) = 0$:
\beq
	\hspace{-6mm}
  \lambda_0(\chi) = -\frac{\Gamma t}{2}
  \left[
    1 - 
    \sqrt{
      1 + \frac{4}{\Gamma^2}\sum_\pm \gamma^\pm \Gamma^\pm(e^{\mp i\chi}-1)
    }
  \right],
\eeq 
with $\Gamma= \gamma^+ + \Gamma^+ + \gamma^- + \Gamma^-$.  
From this expression it is clear that, in the long-time limit, intra-well distances are irrelevant implying that $\ew{x^k}_c \approx a^k \ew{n^k}_c$. The reason is that, in this limit, the extension of the probability distribution function is much larger than a single well. 
Moreover, combining \eq{cums} and \eq{Flong}, one finds that the cumulants are then simply linear in time, as they should be. 
The first two read
\beq
  \ew{x(t)}^\mathrm{asym}_c &=& \frac{\gamma^+ \Gamma^+ - \gamma^- \Gamma^-}{\Gamma}t
  ;
  \\ 
  \ew{x^2(t)}^\mathrm{asym}_c &=&
  \frac{
   \Gamma^2 ( \gamma^+ \Gamma^+ + \gamma^- \Gamma^-)
   -2 (\gamma^+ \Gamma^+ - \gamma^- \Gamma^-)^2
  }{\Gamma^3}t
  .
  \nonumber\\
  \label{x2asym}
\eeq
The latter gives directly the long-time diffusion coefficient 
\beq
  D \equiv
  \lim_{t\to \infty} \frac{1}{2}\frac{d}{dt} \ew{x^2(t)}_c
  =
  \frac{1}{2t}\ew{x^2(t)}^\mathrm{asym}_c
  \label{Dxasym}
  .
\eeq

\section{Rates \label{SEC:RATES}}

The 2SW model has six parameters: the four rates, $\gamma^\pm$ and $\Gamma^\pm$; and the two positions $x_{L/R}$.
The inter-well rates we choose as twice the Kramers' rates 
\beq
  \Gamma^\pm = 2 r_{K\pm}
  \label{interwellrates}
  ,
\eeq 
where $r_{K\pm}$ are the Kramers' rates for passage out of the well to the right ($+$) and left ($-$)\cite{Hanggi1990,Siler10}. Explicit forms are given in \eq{APP:Kramers}.
The factor 2 arises because, if the particle is localized in the right half of the well, the average time it will take to hit the right edge will be half that as if the particle was distributed over the whole well. 
The remaining parameters we fix by comparing the initial behaviour with that of the SE, \eq{SE}.

If we start the SE equation with the particle localized at a point $x_\alpha$ and let the system evolve for a short time, then, to linear order in time $t$, the mean position of the particle evolves as a deterministic particle subject to a constant force $f(x_\alpha) = -u'(x_\alpha)$.  In contrast, the second cumulant evolves as if the particle experiences free diffusion.
Evolution from $x_{L/R}$ therefore gives the first two cumulants
\beq
  \ew{x} &\sim& x_{L/R} + f(x_{L/R}){\scriptstyle \frac{D_0}{kT}} t 
  ;\quad\ew{x^2}_c \sim 2 D_0 t
  .
  \label{FPlinear}
\eeq
Equating these results with \eq{TSQlinear} and rearranging gives
\beq
  \gamma^+ = \gamma^- \equiv \frac{1}{2}\gamma &=& \frac{2 D_0}{(x_R - x_L)^2}
  \nonumber\\
  (x_R-x_L) f(x_L) &=& 2 kT
  \label{feqL}
  \nonumber\\
  (x_R-x_L) f(x_R) &=& -2 kT
  .
  \label{feqR}
\eeq
Equations (\ref{feqL})
have to be solved numerically for $x_L$ and $x_R$, which then give the rates $\gamma^\pm$.

With these choices, since $\gamma^+ = \gamma^-$, we obtain for the plateau values of \eq{plat1} the values 
\beq
  \ew{x}^\mathrm{plat}_c = \frac{1}{2}\rb{x_L + x_R}
  ;\quad
  \ew{x^2}^\mathrm{plat}_c =\frac{1}{4}\rb{x_L-x_R}^2
  ;
  \nonumber \\
  \ew{x^3}^\mathrm{plat}_c= 0
  ;\quad
  \ew{x^4}^\mathrm{plat}_c = -\frac{1}{8}\rb{x_L-x_R}^4
  .
  ~~~~
  \label{plat2}
\eeq
Furthermore, in the limit $\Gamma^\pm/\gamma \gg 1$, the asymptotic cumulants read:
\beq
  \ew{x^k(t)}^\mathrm{asym}_c  \approx \frac{a^k}{2} \left[ \Gamma^+ + (-1)^k \Gamma^-\right]t
  ,
\eeq
which are identical with those of the one-state-per-well model (see Appendix \ref{APP:1S/W})

\section{Results \label{SEC:RESULTS}}

\begin{figure}[tb]
  \includegraphics[width=0.99\columnwidth,clip]{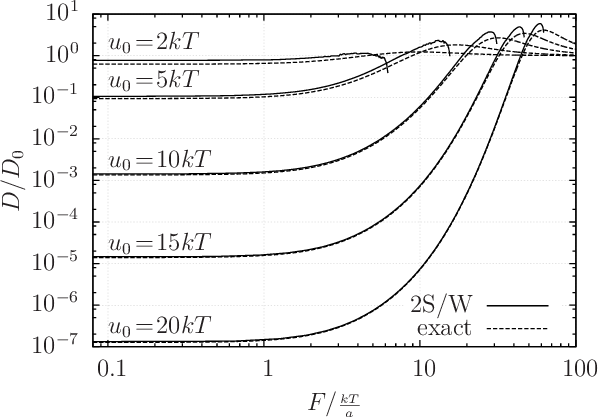}
  \caption{ 
    Comparison of diffusion coefficient $D\!\!=\!\!\lim_{t\to \infty} \frac{1}{2}\frac{d}{dt} \ew{x^2(t)}_c$ of \eq{Dxasym} with the exact result of Ref.~\cite{Reimann01}. 
    The results of the 2SW model agree well with the exact ones in the regime of deep valleys, i.e. $u_0 \gg kT$ and $F \ll F_c$, with $F_c$ the force for which potential valleys disappear.
    \label{FIG:D}
  }
\end{figure}

Before discussing the short-time behaviour of the system, we begin by demonstrating that our 2SW model recovers the known long-time behaviour. In \fig{FIG:D} we compare the diffusion coefficient $D$ from the 2SW model, \eq{Dxasym}, with the exact result of Reimann {\it et al.} \cite{Reimann01}.  In accordance with expectations, our model works well provided that the valleys of the potential are sufficiently deep, $\Delta u \gg kT$.  This condition can be satisfied by ensuring that $u_0\gg kT$ and that the applied force is less that the value of the force $F_c\!=\!u_0\pi/a$ at which the minima of the potential disappear.

\begin{figure}[tb]
  \includegraphics[width=0.99\columnwidth,clip]{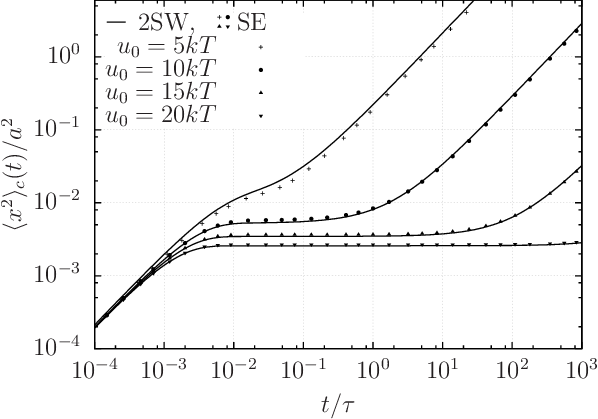}
  \caption{ 
    The mean squared displacement $\ew{x^2(t)}_c$ for different amplitudes $u_0$ with no applied force ($F=0$). The lines show the result from the 2SW model, and the symbols those from numerical integration of the SE. For the parameters shown here, the 2SW model reproduces the essential features of the short-time dynamics.  The agreement with the numerically exact results increases with increasing well-depth (increasing $u_0$).
    \label{FIG:C2t}
  }
\end{figure}

\begin{figure}[tb]
  \includegraphics[width=0.98\columnwidth,clip]{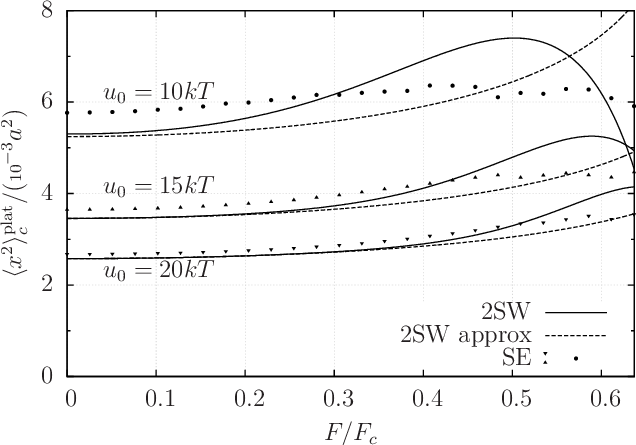}
  \caption{
    Plateau height $\ew{x^2}_c^\mathrm{plat}$ as a function of driving force, $F$, calculated in three different ways.
	 We have defined the plateau height as the value of $\ew{x^2}_c$ at the time for which $d(\ln \ew{x^2}_c)/d(\ln t)$ reaches its minimum value.
    The solid curves show the plateau heights obtained from the 2SW theory, the symbols from integrating the SE and the dashed lines show the approximation of \eq{plat2}.
    The 2SW method underestimates the plateau heights slightly, becoming better as $u_0/kT$ becomes larger.
    Only when the two 2SW results agree with one another does the cumulant really show a well-defined plateau.
    \label{FIG:plateau}
 }
\end{figure}

Turning now to the short-time dynamics, we consider first the second cumulant $\ew{x^2(t)}_c$ as a function of time in \fig{FIG:C2t}, where time is measured in units of $\tau=\frac{a^2}{D_0}$.  Three distinct behaviours are seen.
At the shortest times, the behaviour is essentially free diffusion with the second cumulant growing linearly. This is the behaviour which was fitted in our 2SW model to obtain Eqs.~(\ref{feqR}).
As the particle begins to feel the influence of the wells, a plateau develops in the mean squared displacement.  The duration of this plateau increases the higher the potential barrier between wells.
This behaviour is well reproduced by the 2SW model not just qualitatively, but quantitatively. 
We note that this behaviour was not explicitly put into the model; rather, it emerges as a prediction of it.
\fig{FIG:plateau} shows the height of the plateau as a function of the applied force $F$.  The 2SW approach works better the more pronounced the potential barriers, but even for $u_0=5kT$, the 2SW theory and numerics agree quite well. 
\fig{FIG:plateau} shows two different results for the plateau height extracted from the 2SW model; one based on the full solution and one given by the approximate value of \eq{plat2}.   The coincidence of these results is a good indication that the mean-square displacement undergoes a genuine plateau, rather than something more like an inflexion point, as is the case for e.~g. $u_0/kT = 5$ in \fig{FIG:C2t}.
Not shown here, but the first cumulant $\ew{x}_c$ shows similar plateau behaviour, the height and duration of which is also well reproduced by the 2SW model.

\begin{figure}[t]
  \includegraphics[width=0.98\columnwidth,clip]{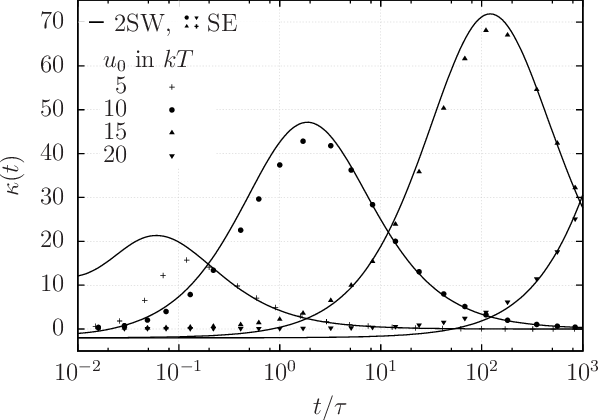}
  \caption{
    The kurtosis $\kappa(t)$ of \eq{kappa} as a function of time for the same parameters as in \fig{FIG:C2t}.   The dominant feature is a peak which occurs around the cross-over time.
    As before, good agreement between 2SW model and numerics is obtained for $u_0 \gg kT$.
    \label{FIG:kappat}
 }
\end{figure}

\begin{figure}[t]
  \includegraphics[width=0.98\columnwidth,clip]{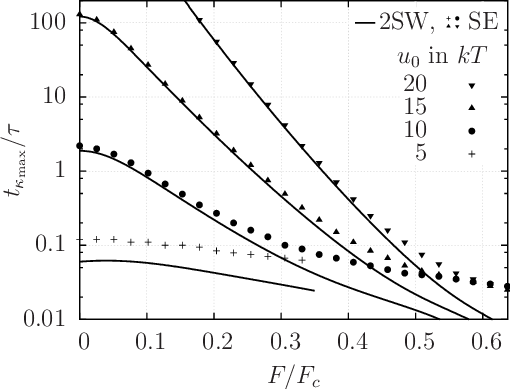}
  \caption{
    The time of the kurtosis peak as a function of driving force $F$ for different values of $u_0$. Again, the agreement is good for deep valleys.
    \label{FIG:kappamax}
 }
\end{figure}

At longer times, the particle breaks out of the confinement of its initial well, and starts diffusing through the potential.  The second cumulant then starts increasing linearly again, this time with the diffusion constant of \eq{Dxasym} and \fig{FIG:D}.  The time at which the plateau goes over into diffusive behaviour is the  cross-over time.
During the plateau and through the cross-over, the probability density slowly reshapes.  This can be seen in \fig{FIG:kappat}, which shows the kurtosis, \eq{kappa}, as a function of time.  By comparing \fig{FIG:C2t} and \fig{FIG:kappat} one can see that the time of the maximum deviation from Gaussian e.g.\ the maximum of $\kappa(t)$ can be used as an approximation to the end of the plateau, as proposed in Ref.~\cite{Vorse07}.
It is remarkable that the dependence of this crossover time on $u_0$ and $F$, shown in \fig{FIG:kappamax}, is given in common quantitatively.

\section{Conclusions \label{SEC:CONCS}}
In this paper we have investigated the short-time dynamics of a particle driven through a modulated potential using a Markovian master equation approach based on a discretization of the spatial axis into two states per well. The remaining input into the theory are the transition rates, which we have chosen by using Kramers' rates (inter-well transitions) and by comparison
with the true short-time dynamics from the continuous (SE) approach (intra-well transitions).

A major advantage of our discretized two-state model is that relevant transport quantities, such as the mean-squared displacement, can be calculated analytically.
In particular, we have derived explicit expressions for the heights of the plateaus in the mean-squared displacement at intermediate times, i.e., within the subdiffusive regime,
and for the kurtosis, which signals the crossover to the diffusive long-time regime. 
The results from the two-state model are in excellent agreement with those obtained from numerical solution of the SE with and without external drive, provided that $\Delta u /kT \gg 1$ and $F\lesssim F_c$. 
This latter requirement is consistent with the expectation that (only) in this limit, the valleys in our continuous model can be approximated
by almost-isolated wells. In the same range of parameters, the diffusion coefficient $D$ (derived from the second cumulant), is found to be fully consistent with 
previous analytical results of Reimann {\it et al.} \cite{Reimann01}.
It is precisely the high-barrier case, which is relevant in many experimental realizations
of colloids in modulated potentials (see e.g.\ Ref.\ \cite{Dalle11}), as well as in the more general context of modeling systems exhibiting cage effects (such as undercooled liquids \cite{Chaud08}
and dense colloidal suspensions \cite{Vorse07}). For such systems our ``minimal'' model for short-time dynamics could be particularly useful.

We close this paper with some remarks on possible extensions of our approach and its relation to other models. 
To start with, we have found that discrepancies between the master equation and the SE approach arise as $\Delta u /kT$ decreases.  In this case,
quantitative agreement can be improved by modifying the manner in which the rates are calculated.  Firstly,  expanding the full generating function to linear order in time yields
\beq
  \ew{x(t)} &\sim& x_{L/R} \pm  \gamma^\pm (x_R - x_L) t 
    \mp \Gamma^\mp (a  + x_L - x_R)t
  \nonumber\\
  \ew{x^2(t)}_c &\sim&  \gamma^\pm (x_R - x_L)^2 t 
    + \Gamma^\mp(a  + x_L - x_R)^2 t.
\eeq
These values can be equated to \eq{FPlinear} to match better with the SE. Secondly, the inter-well rates, \eq{interwellrates}, can be improved by using higher-order corrections to Kramers' rate \cite{Risken}.  These latter corrections are particularly important when the periodic contribution to the potential becomes steeper than the sinusoid considered here.

A further point concerns the performance of our master equation approach for the prediction of higher-order cumulants. Whereas the complete behaviour of the first two cumulants is well reproduced by our 2SW model, the agreement for the higher-order cumulants is not as good, and although the broad features of these $k>2$ are reproduced, some fine details at very short times are not captured.  To improve these higher cumulants it seems to be promising to augment the model with more states per well, an extension which is conceptually straightforward.

Finally, we note that there is an connection between our model and other models describing subdiffusion.  By tracing over the internal degrees of freedom, a nonMarkovian master equation for the well-index $n$ alone can be derived, involving a memory kernel which differs from a delta function (characterizing the Markovian case). Specifically, in the 2SW model, the kernel in Laplace space has a single simple pole at $-\Gamma$, which describes the rate at which the intra-well degrees of freedom relax. 
It is well established \cite{Haus1987} that such non-Markovian kernels are intimately related \cite{Bedeaux71,Krenke73} to the waiting-time distribution in continuous-time random walks \cite{Montroll65}, such that connexion can also be made with these latter methods.

\begin{acknowledgments}
This work was supported by the Deutsche Forschungsgemeinschaft through SFB-910.
\end{acknowledgments}

\appendix
\section{One-state-per-well model \label{APP:1S/W}}

The simplest approach is to split the potential into a number of wells (labelled $n = \ldots,-2,-1,0,1,2,\ldots$ ) defined between subsequent maxima, and write down a rate equation for the probability $p^{(n)}(t)$ of finding the particle in well $n$ at time t. The position of the particles is given as $x=n a$.  The rate equation looks like
\beq
  \dot{p}^{(n)} = -(W^+ +W^-) p^{(n)} + W^+ p^{(n-1)} +  W^- p^{(n+1)}~~~~~~
  \label{1SPW_RE}
  ,
\eeq
where $W^{\pm}$ are rates.  We define the MGF and CGF via
\beq
  e^{\mathcal{F}(\chi,t)}=\mathcal{G}(\chi,t) = \sum_n e^{i n \chi} p^{(n)}
  ,
\eeq
such that the $k$th derivative
\beq
  \frac{\partial^k}{\partial(i\chi)^k}\mathcal{F}(\chi,t)
  |_0 = \ew{n^k(t)}_\mathrm{c}
\eeq
is the $k$th cumulant of the well number. The corresponding position cumulants are simply: $\ew{x^k(t)}_c = a^k \ew{n^k(t)}_c$.  Fourier Transform of \eq{1SPW_RE} gives
\beq
   \dot{\mathcal{G}}(\chi,t)  &=& W(\chi) \mathcal{G}(\chi,t)
\eeq
with
$
  W(\chi) = \sum_\pm W^\pm (e^{\pm i \chi}-1)
$.  Solving this equation assuming that we start localised in the $n=0$ well, gives the CGF
\beq
  \mathcal{F}(\chi,t) =  W(\chi) t
  .
\eeq
Differentiating, we obtain the cumulants
\beq
  \ew{x^k(t)}_c = a^k \left[ W^+ + (-1)^k W^-\right]t{}
  .
  \label{1SPW_cums}
\eeq
The cumulants of this model are therefore linear in time for all time, and there can be no caging effects in this one-state-per-well model. All odd cumulants are equal. All even cumulants are equal.

Good agreement with the asymptotic solutions \cite{Reimann01} is obtained in the regime $\Delta u/kT\gg1$ with the rates chosen according to Kramers rule,  $W^\pm = r_{K\pm}$ with \cite{Hanggi1990,Siler10}
\begin{align}
	r_{K+}&=\frac{D_0}{2\pi kT}\sqrt{|u''(x_\mathrm{min})||u''(x_c)|}
	\;e^{-(u(x_c)-u(x_\mathrm{min}))/kT}
	\nonumber
	\\
	r_{K-}&=\frac{D_0}{2\pi kT}\sqrt{|u''(x_\mathrm{min})||u''(x_a)|}
	\;e^{-(u(x_a)-u(x_\mathrm{min}))/kT}
	\label{APP:Kramers}
  ,
\end{align}
where $x_\mathrm{min}$, $x_a$, $x_c$ are the positions of a potential minimum and the maxima to its left and right, respectively.

\section{ CGF \label{APP:CGF}}
We demonstrate here that \eq{CGFx} gives the position cumulants as stated in \eq{cums}. Consider the moment generating function corresponding to \eq{CGFx}:
\beq
  \mathcal{G}(\chi;t) =  e^{\mathcal{F}(\chi;t)}
  =\braa{\widetilde{\phi}_0} e^{i \chi \Delta}  e^{ \mathcal{W}(\chi) t} \kett{\rho^{(0)}(0)} 
  .
  \label{MGFx}
\eeq
Taking the $k$th-derivative, we find
\begin{widetext}
\beq
   a^k\left.
  \frac{\partial^k}{\partial (i\chi)^k}  
  \mathcal{G}(\chi;t)
  \right|_{\chi=0}
  &=&
   a^k\left.
  \frac{\partial^k}{\partial (i\chi)^k}  
  \braa{\widetilde{\phi}_0} e^{i \chi \Delta}  \kett{\rho(\chi;t)} 
  \right|_{\chi=0}
  =
   a^k\left.
  \frac{\partial^k}{\partial (i\chi)^k}  
  \sum_n
  \braa{\widetilde{\phi}_0} 
  \sum e^{i \chi (\Delta +n)} \kett{\rho^{(n)}(t)} 
  \right|_{\chi=0}
  \nonumber\\
  &=&
  \sum_n
  \braa{\widetilde{\phi}_0} 
   (\Delta +n)^k \kett{\rho^{(n)}(t)} 
  =
  \sum_n (n a+ x_L)^k p^{(n)}_L 
  + (n a + x_R)^k p^{(n)}_R
  = \ew{x^k}_m
  ,
\eeq
which is the $k$th moment of the spatial distribution. The standard relation between MGF and CGF completes the proof.
\end{widetext}




\begin{thebibliography}{00}
\bibitem{FN}  Although there is no physical barrier separating the two states within a well, the description of the movement of the particle between these states in terms of rates makes sense, just as the diffusion of a particle on a flat surface can be modelled by the familiar random walk.
\bibitem{FNtracevec}
  This can be seen by considering, e.g., the scalar product 
  $
  \eww{ \phi_0 | \rho(\chi=0;t)}
  =
  \rb{1,1}^T 
  \cdot \rb{\sum_n p_L^{(n)},\sum_n p_R^{(n)}}
  =\sum_n \rb{ p_L^{(n)} + p_R^{(n)}}
  =1
  $.
\end{thebibliography}
\end{document}